**Direct Measurement of the Effective Electronic Temperature in Organic Semiconductors**

Anton Kompatscher, Martijn Kemerink*

Institute for Molecular Systems Engineering and Advanced Materials, Heidelberg University, Im Neuenheimer Feld 225, 69120 Heidelberg, Germany

*corresponding author; email: martijn.kemerink@uni-heidelberg.de

Abstract:

Organic semiconductors show complex phenomena due to their high energetic disorder. A striking example is the possibility of an increased effective temperature $T_{eff}$ of the charge carrier distribution relative to the lattice temperature, which results from the slow charge carrier relaxation after excitation, either by high electric field or photon absorption. The increased effective temperature has been linked to conductivity enhancements and performance increases in actual devices, but a direct observation has been lacking. Here, we utilize nanoscopic tree-terminal devices to measure the Seebeck voltage arising in a doped organic polymer semiconductor due to a field-driven enhancement of the effective electronic temperature, providing direct proof of the existence of $T_{eff}$. The results agree quantitatively with numerical predictions by a kinetic Monte Carlo model. The findings not only provide fundamental understanding but also indicate an avenue towards low-loss thermoelectric devices.

While it is widely accepted that disorder is a crucial characteristic of virtually all organic semiconductors, there is still debate about its exact influence on charge transport. [1–4] This becomes apparent in the discussion about field behavior. Initially, the effect of conductivity increasing with electric field strength was described in terms of Pool-Frenkel behavior of the type $\sigma \propto \exp(\sqrt{F})$. [5] It is argued, however, that this description is phenomenological at best, specifically at high fields and alternative approaches have been proposed. [1,6–8] A common and physically elegant approach focuses on the role of slowly relaxing carriers, which allow the energy distribution to be driven out of equilibrium by modest fields of order $10^7$ V/m, which are not uncommon in typical organic thin film devices. [9] The effect of the applied field is then described as a modified, increased, temperature of the charge carrier distribution. This concept is illustrated for electrons in Figure 1, but holds, mutatis mutandis, for holes as well. As the charges hop under influence of an electric field, they gain additional energy due to the drop in potential. If carrier relaxation is sufficiently slow, the distribution of occupied states shifts upwards as if having an elevated temperature, called the effective temperature ($T_{eff}$), relative to the unchanged lattice temperature. From simulations it was phenomenologically described as:

$$T_{eff} = \left[ T_{lat}^\beta + \left( \gamma \frac{e|F|\alpha}{k_b} \right)^\beta \right]^{\frac{1}{\beta}} \qquad (1)$$

with the lattice temperature $T_{lat}$, the electric field strength $F$, the localization length α and the constants β > 1 and γ = 0.3 − 0.9. The constants $e$ and $k_b$ represent the elementary charge and the Boltzmann constant, respectively. [10–13] Conductivity is then obtained by replacing the lattice temperature with effective temperature in the underlying gaussian disorder model (GDM), i.e. $\sigma(T_{lat}, F) = \sigma(T_{eff}(T_{lat}, F), F = 0)$. [1] While this concept is well-tested in simulations of single materials, the exact expression lacks a physical justification. In addition, on basis of master equation simulations, the concept has been argued to be inapplicable in host-guest systems. [14] Surprisingly little experimental work is devoted to putting the concept on more solid footing, especially for organic semiconductors. [7] In part, this may be due to the combination of the (large) energetic disorder and exciton binding energy that preclude the use of conventional optical approaches to measure electronic temperatures. [9,15]

The previous experiments known to us can be separated into two approaches, which have both been pursued by Nebel et al. in amorphous hydrogenated silicon. The first directly measures temperature and field scaling of conductivity and checks for equivalence when applying the effective temperature model. [16] The second relies on photoconductivity experiments performed in p-i-n junctions. [17] Here, the equivalent scaling with field and temperature of the initial current decay after photoexcitation and also the drift mobility, is analyzed. For Nebel et al. all approaches yield good fits with slightly varying localization lengths between 5- to 10 Å. Follow-up photoconductivity experiments in the same material are mixed, with some studies finding reasonable agreement [18,19], while Gu et al. [20] cannot fully describe their results in terms of the model, attributing this to improved photoexcitation conditions. Later studies on amorphous carbon compounds also utilize modified versions of the effective temperature concept treating γα as a field-dependent variable. [21] These existing experiments are mostly indirect in the sense that they demonstrate only the equivalence between temperature- and field-scaling of conductivity or current decay when expressing field in terms of temperature as per the equation above, which formally does not prove the existence of a (more or less) thermal distribution of charge carriers characterized by an enhanced temperature. [16]

In line with the above, the possibility of a raised effective temperature is rarely considered in device design, despite -potentially- substantial implications. As recently demonstrated, the efficiency of organic solar cells may be significantly improved by their ability to utilize a similar distribution shift due to the excess energy of incoming photons. [15,22] Here, the idea is that charges driven out of equilibrium generate an (additional) Seebeck voltage, enhancing device performance. In case it would be possible to couple general excess energy, for instance in the form of stray electric fields, to the charge carrier distribution only, rather than to the full device including the lattice, highly efficient field-driven thermoelectric devices could be realized. [23,24] The reason is that raising only the charge carrier temperature leads to a device that is not, or less, affected by the lattice thermal conductivity, which represents a significant loss channel in current thermoelectric devices. This can be illustrated by the figure of merit for thermoelectrics $ZT = S^2 \sigma_e T/(\kappa_e + \kappa_l)$. [25,26] Using the Wiedemann-Franz law as lowest order approximation for the electronic part of the thermal conductivity $\kappa_e = L_0 T \sigma_e$ with $\sigma_e$ the electrical conductivity and $L_0 = \frac{\pi^2}{3}\left(\frac{k_B}{q}\right)^2 \cong 2.44 \times 10^{-8}$ V$^2$K$^{-2}$ the Lorenz number, gives, for negligible lattice thermal conductivity $\kappa_l$, $ZT = S^2/L_0$. [27] For reasonable values of the Seebeck coefficient $S > 200$ µV/K, one then gets $ZT > 1.6$, which is well beyond what is currently available around room temperature.

Here, we aim to directly measure the effective temperature through an electric field-driven Seebeck voltage to directly prove the existence of a locally thermal but globally nonequilibrium distribution of charge carriers in the system. To maximize the collected signal and to minimize the voltage amplitudes, we opted for structures at the nanoscale that give significant electric field strengths at low voltages. Specifically, this reduces the influence of any higher harmonic signals from, e.g., function generators of amplifiers that are unavoidably present in the driving signal and that might overwhelm the signal of interest.

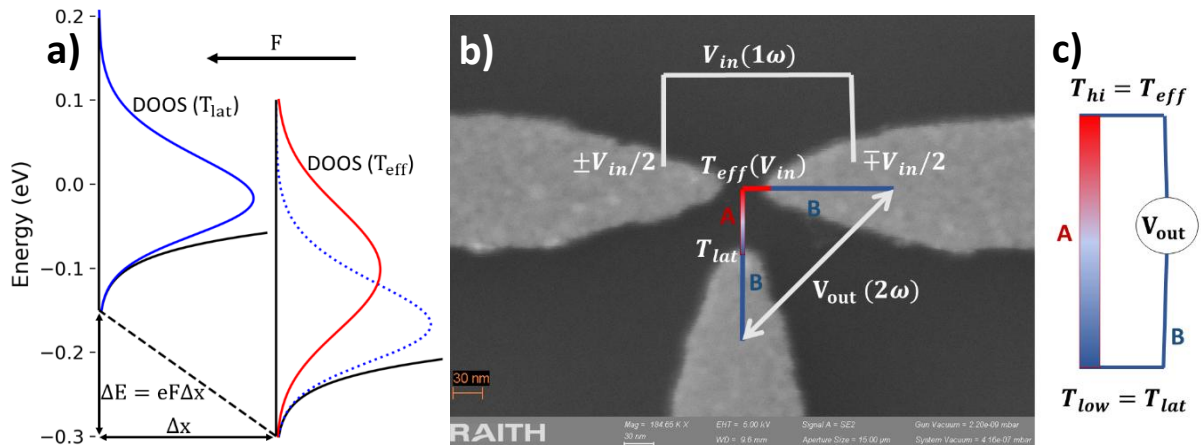

Figure 1: a) Schematic illustration of the shift of the electron density of occupied states (DOOS, colored lines) under the application of an electric field $F$. The density of states (DOS, grey and black lines), shifts with the potential as charges travel to a site located at a distance Δx. Due to the slow carrier relaxation, the zero-field lattice temperature distribution (blue) is replaced by the effective temperature distribution (red). b) Schematic depiction of the experiment, showing the input and output circuit overlaid on an electron microscopy picture of the center tips. For the actual experiment, the entire structure is covered in F4TCNQ-doped P3HT. The application of a voltage $V_{in}$ at frequency ω leads to effective temperature $\propto V_{in}^2$ and hence a Seebeck voltage $V_{out}$ at 2ω. c) The equivalent Seebeck circuit with materials A and B representing the organic and gold layers, respectively.

As visible in Figure 1, the experiment is set up with tree electrodes: two opposing input electrodes and one perpendicular measurement electrode. In order to improve yield and identify possible trends, we manufactured 30 of such structures with varying distances per (~1.2×1.2 cm$^2$) substrate. The experiment was realized on n-doped silicon substrates with a thermally grown 200 nm silicon dioxide cover layer. The nanostructures were manufactured by electron beam lithography in a Raith e-line pioneer system. The liftoff layer was made from PMMA 950K. The structures were developed in MIBK (4-methylpentan-2-on) and isopropanol; the liftoff was performed with NEP (n-ethyl-2-pyrrolidone). The metal layers of 5 nm of chromium and 10 nm of gold were evaporated from an efm3T electron beam evaporator. The base pressure during evaporation was $\sim 10^{-7}$ mbar. The organic films where spin coated on the lithographic structures in a glovebox at a < 3 ppm $O_2$, $H_2O$ nitrogen atmosphere. For bulk doping of regio-regular poly(3-hexylthiophene-2,5-diyl) (rr-P3HT) and 2,3,5,6-tetrafluor-7,7,8,8-tetracyanochindimethane ($F_4$TCNQ) were both dissolved in orthodichlorobenzene (ODCB) at a P3HT concentration of 20 mg/ml. All given doping ratios are molecular ratios. Depending on substrate size $\sim 60\ \mu L$ of solution was statically spin coated at 1000 rpm for 60 seconds and dried at 3000rpm for 25 seconds. The resulting films typically had thicknesses between 60 and 100 nm. To test the behavior at high doping, sequentially doped samples were fabricated in two spincoating steps. The first consisted of a pure P3HT film as described above. In the second step 5 mg/ml F4TCNQ were dissolved in a 4:1 mixture of tetrahydrofuran to dichloromethane of which then $\sim 80\ \mu L$ was spincoated dynamically onto the P3HT film as described by Zuo et al [28]. While this is a hole transporting system, the underlying principles are equally applicable to electrons.

Applying an AC voltage signal between the inputs heats the carrier distribution between them at twice the input frequency due to the direction invariance. The measurement electrode is situated outside the high field region and charge carriers in its vicinity are therefore at a lower (effective) temperature. We thus expect to measure an oscillating Seebeck voltage $\Delta V \propto S(T_{eff} - T_{lat})$ at twice the input frequency between the input ground and output. [29] Utilizing a lock-in amplifier, we can filter for these frequencies and exclude currents due to the applied potential and unintentional asymmetry in the device as these will occur at 1ω. An unwanted contribution we cannot filter via frequency is ohmic heating of the full area around the driving tips due to the input power $P = V^2/R$. As it similarly scales with the square of the applied voltage, it will also oscillate at the same frequency as the effective temperature, but it can be suppressed by using high-resistance samples. We will thus have to analyze which Seebeck voltages stem from lattice temperature differences and which are actually caused by charge carrier temperature differences.

We first discuss the raw data in order to identify the signal of interest. Representative raw data of our experiments can be found in Figure 2 which depicts the first and second harmonic signal measured at the output electrode for a bulk doped sample. Between 10 and 1000 Hz a steep decline of the signal is visible which we attribute to the RC-cutoff of our setup. Given the high sample resistance $R$ of ~100 MΩ and the ~40pF input capacitance $C$ of our measurement device, we arrive at a cutoff frequency $f_c = 1/2\pi RC$ of ~40 Hz which agrees well with the observed drop in $V_{out}$ and matches the frequency at which current and voltage measurements diverge in figure 2b. This is further supported by the differing cutoffs in the sequentially doped sample visible in figure S1 (a,b) in the supporting information (SI) with lower resistance.

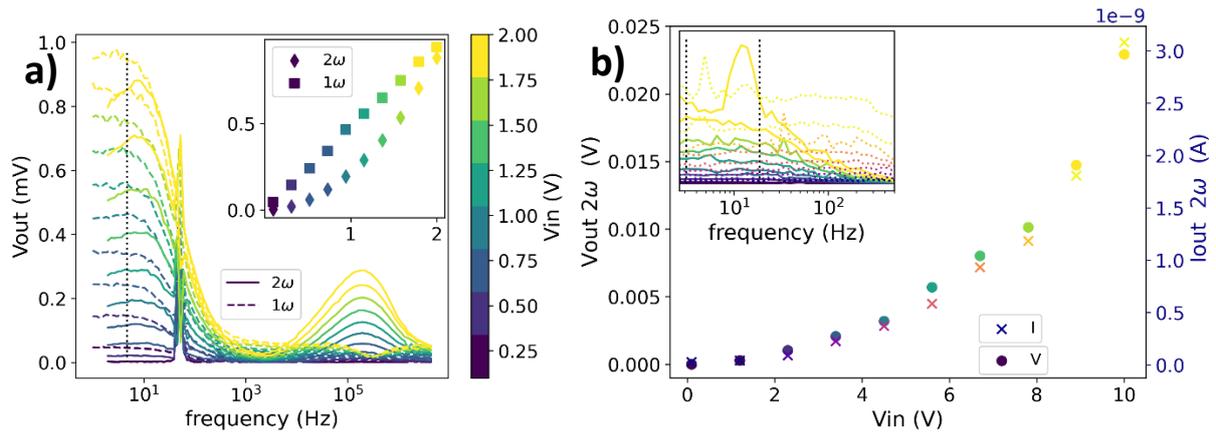

Figure 2: a) Raw data of first and second harmonic response measured on a P3HT film bulk doped with 0.1% molecular ratio F4TCNQ. The input tips are separated by a 40 nm gap and the measurement tip is 60 nm from the center line. The inset shows a cut through the data at a frequency of 4.75 Hz indicated by the dotted line. b) Raw data of voltage and current output on a 1% bulk doped film averaged between 3 and 19 Hz. Here, the input and measurement distances were 15 and 33 nm. Voltage and current follow the same trend in the indicated frequency range.

The peak at roughly 0.1 MHz is, in contrast to the signal below 100 Hz, also visible when the structure is not coated with an organic layer, see SI figure S3. Therefore, we attribute it to a resonance of the structure itself and do not analyze it further. Thus, the signal of interest is situated in the low frequency regime below 100 Hz. The inset shows the scaling of first and second harmonic signals at a sufficiently low frequency of 4.75 Hz, representative of the region of interest. The sample shows a clear trend of the first harmonic signal increasing linearly. Being a first harmonic signal, this cannot be attributed to ohmic heating or effective temperature, and we attribute it to a minor asymmetry in the device, such that the symmetric drive, c.f. figure 1b, does not average out to exactly zero. The relevant second harmonic signal, in contrast, increases nonlinearly as expected from the effective temperature formula Eq. 1. Being super linear, the increase cannot be attributed to a higher harmonic in the driving signal in combination with the same geometric asymmetry as for the first harmonic signal. We can further exclude that second harmonic generation in the signal source is causing this, as the latter is measured to be several orders of magnitude smaller than the first harmonic signal, while in the device the first and second harmonics are of the same magnitude (SI section 2).

As mentioned in the introduction, a possible spurious cause for a second harmonic signal is the local ohmic heating of the sample due to the applied driving voltage. As this would create a very similar, but strictly quadratic, nonlinear signal due to the dissipated power being voltage squared over resistance, we have to exclude it. To this end we firstly calculated the expected ohmic heating in our devices by finite element calculations, which are presented in the SI section 3. At low conductance, as expected for our low concentration bulk doped samples, the expected ohmic heating falls below 1 K, which is far too low to explain the measured signals. Samples with higher doping can theoretically develop significant ohmic heating and are therefore not discussed in the main text. More details can be found in the supporting information, sections 1 and 3. For comparison with the functional dependence expected for the effective temperature model, we have also included quadratic fits in our graphs. Where there is a difference between the two, the former gives the better description of the data.

Having identified the signal of interest as the low frequency second harmonic signal, we will now perform a quantitative analysis. We noted both a relatively large inter-sample variability and noise-like, i.e. stochastic, variations in individual measurements that we both attribute to the limited probing volume, making measurements sensitive to sample-to-sample variations and (field- or temperature-driven) variations in the position of individual dopants or even polymer segments, as further discussed in SI section 4. We will focus on one representative measurement here.

Focusing on low conductivity samples brings the need to account for the high sample resistances relative to the measuring instrument. As the used lock-in amplifier has an input impedance of 10 MΩ, the actual signal is higher than measured due to the voltage division between the signal input resistance and the sample resistance and the measured voltages were corrected accordingly. More details on the used correction schemes can be found in the supporting information section 1.

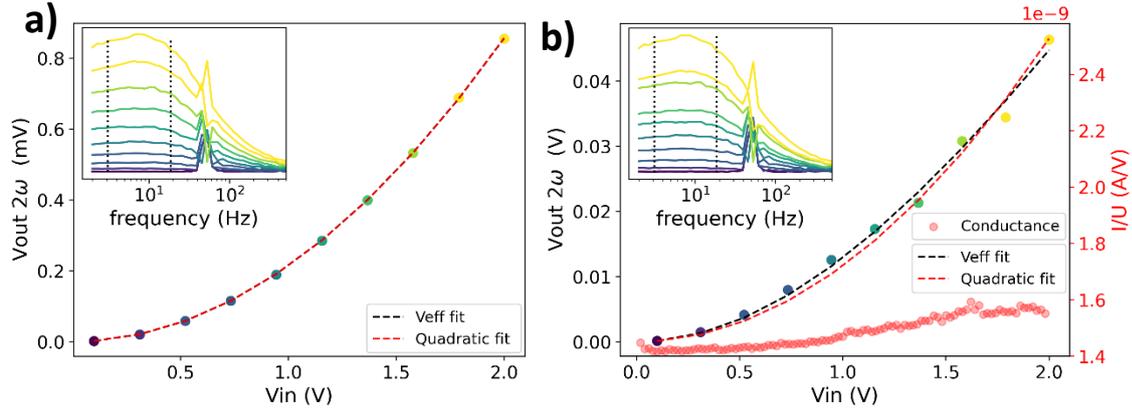

Figure 3: Cuts through the measurement fitted with equation ( 2 ) and a quadratic function for comparison. The P3HT films were bulk doped with 0.1% F4TCNQ. Each point is the average between all measurements in the frequency range indicated by the black dotted lines. Plot b) shows the same second harmonic voltage after correction for the resistance mismatch between sample and measurement device. The conductance data (red dots) utilized to calculate this correction is also displayed. The input and measurement distances were 35 and 60 nm, respectively. Corresponding IV-curves measured between the input tips are shown and discussed in SI section 7.

We are now ready to analyze the (corrected, low-frequency) data as Seebeck voltages, generated due to the temperature difference between the 'hot' high-field region between the driving tips and the 'cold' probe tip. For the former, we use equation ( 1 ), assuming $\beta = 2$ and for the latter the lattice temperature. In doing so we assumed that the applied voltage drops uniformly over the input tip separation $F \approx \frac{Vin}{d}$. Via numerical simulations we verified that the effect of field inhomogeneities would only lead to a correction of factor unity which is small compared to other variations between measurements and corrections (see Si section 5). Hence, we arrive at the fitting expression

$$V_{eff} = S\left(\left(T_0^2 + \left(0.89 \frac{V_{in}}{d} \frac{\alpha e}{k_b}\right)^2\right)^{\frac{1}{2}} - T_0\right) \quad (2)$$

For the uncorrected data presented in Figure 3a this leads to a fitted localization length of $0.14 \pm 1.9$ Å and a fitted Seebeck coefficient of $7 \pm 200$ mV/K. After correcting for the resistance mismatch, figure 3b, we arrive at a localization length of $5.3 \pm 1.8$ Å and a Seebeck coefficient of $330 \pm 190$ μV/K. As we expect localization lengths in the several Å to few nm range [30,31] and Seebeck coefficients around a few 100s to 1000 μV/K [28], fitted values after impedance correction are reasonable, especially given the experimental uncertainties. On uncorrected data, the effective temperature fit produces the same shape as a quadratic fit meant to describe ohmic heating. When the impedance mismatch corrections are applied, output voltage becomes more linear and is better matched by the effective temperature model.

Utilizing the fitted Seebeck coefficient, the resistance-corrected data in Figure 3 correspond to an effective temperature about 140 K above the lattice temperature at a field of $5.7 \times 10^7$ V/m. This not only far exceeds the predicted ohmic heating of less than 1 K, it would also lead to immediate sample degradation if present in the device as an actual lattice temperature. [32] To put the number further into perspective, Nebel et al. have analyzed the effective temperature in doped hydrogenated silicon via its conductivity and report estimated effective hole temperatures of 180 K at a lattice temperature of 10 K and a field of $\sim 6 \times 10^7$ V/m. [16] They assumed a localization length of 5 Å for holes, well matching with our 5.25 Å, while suggesting some field dependence of the localization length might exist. Photoconductivity experiments by the same authors place the localization length at 6 Å [17] Given that these are two different materials at different temperatures, this indicates not only a reasonable agreement, but also an (anticipated) universality across disordered systems of various types.

The measurement also excellently agrees with the effective temperatures we predicted in an earlier publication via the use of kinetic Monte-Carlo simulations, as shown in

Figure 4. [23] As the simulation data are obtained for a lattice at room temperature (300 K), a direct comparison with the experimental data is possible. Since the simulation data was created by us before the experiment, it is free from any selection bias. Moreover, and more importantly, the quantitative agreement with experiment is a very strong argument in favor of the physical reality of a field-driven enhancement of the electronic temperature in the true meaning of the word, that is, the density of occupied states forming a thermal distribution characterized by $T_{eff} \geq T_{lat}$, as (still phenomenologically) captured by Eq. 1. Since the simulation uses a framework and parameters that are generic for strongly energetically disordered semiconductors, these findings imply a general validity of the concept.

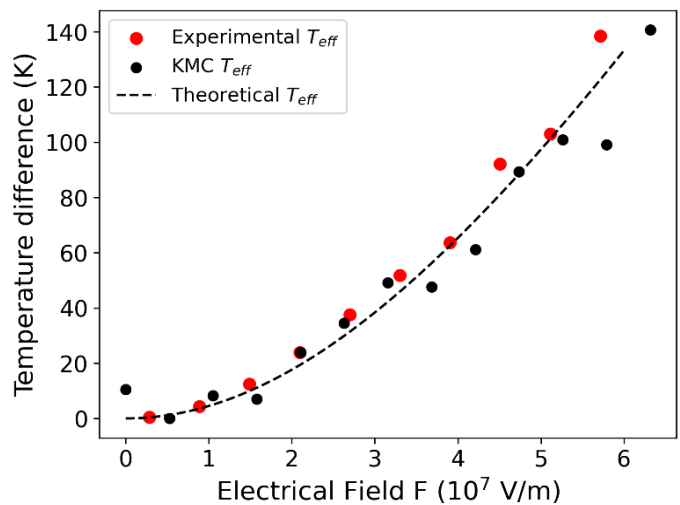

Figure 4: Comparison of the difference between the effective electronic temperature and the lattice temperature, calculated from the experimental data of Figure 3 (red symbols) and from kinetic Monte Carlo simulations from Ref. [23] (black symbols). The experimental data are measured (resistance-corrected) voltages divided by the fitted Seebeck coefficient. The dashed black line is the functional shape of the effective temperature formula with parameters $\alpha = 0.5$ nm, $\beta = 2$ and $\gamma = 0.89$.

In conclusion, we utilized an AC tree-terminal experiment to test the effective temperature concept that predicts an elevated electronic temperature, that is different from the lattice temperature, once the energy scale associated with hopping along the electric field, $eF\alpha$, becomes comparable to or larger than the thermal energy $k_B T$. As expected for periodic heating, we found a second harmonic response that corresponds to the elevated electronic temperature. By frequency-selective amplification and measuring low-doped, high-resistance films, we were able to exclude sample asymmetry and ohmic heating as spurious causes for the signal. Analyzing the measured thermovoltages with the effective temperature formalism first proposed by Marianer and Shklovskii yielded realistic values for the Seebeck coefficient and the localization length. [10] The results give a good match with previous results and an excellent match with theoretical predictions, indicating a universal applicability to strongly energetically disordered materials.

The experimental demonstration that the electronic temperature is largely decoupled from the lattice temperature potentially has far-reaching consequences for understanding thermalization and high-field behavior of existing device designs like solar cells and LEDs. While this knowledge is helpful for understanding, designing and improving existing devices, new concepts seem possible as well. Here, we think less of conventional thermoelectric generators – even if these would greatly benefit from suppressing the losses due to lattice thermal conductivity, this would require to couple the (waste) heat source to the charge carrier distribution only, which is far from trivial. Instead, we imagine Seebeck-based or -enhanced devices for the conversion of light. For instance, our results reinforce previous ideas for hot carrier solar cells that could thus be realistically based on disordered semiconductors. [15,22,33] An alternative approach might be to use plasmonic structures to couple the light field of incoming radiation into the charge carrier distribution, which is then rectified via the Seebeck effect. In all cases, the potentially large temperature differences between effective and lattice temperature, in combination with an enhanced figure of merit $ZT$, enable much larger power conversion efficiencies than conventional thermoelectric generators offer.

Acknowledgements:

We thank Stefan Kauschke and the IMSEAM core facility for technical support. This research is funded by the Deutsche Forschungsgemeinschaft (DFG, German Research Foundation) under Germany`s Excellence Strategy – 2082/1 – 390761711. M.K. thanks the Carl Zeiss Foundation for financial support.

Conflict of interest:

The authors declare no conflict of interest.

Data availability:

Data are available from the authors upon reasonable request.

Supporting Information for:

**Direct Measurement of the Effective Electronic Temperature in Organic Semiconductors**

Anton Kompatscher, Martijn Kemerink*

Institute for Molecular Systems Engineering and Advanced Materials, Heidelberg University, Im Neuenheimer Feld 225, 69120 Heidelberg, Germany

*corresponding author; email: martijn.kemerink@uni-heidelberg.de

# Contents

# 1. Additional samples and corrections:

This section shows a number of additional measurements with varying doping levels. We want to point out the high resistance of most measured films. As the measurement device has an input impedance of 10 MΩ, the actual signal is higher than the measured one due to the voltage division between the lock-in amplifier input resistance and the sample. We can estimate correction factors utilizing the DC resistance measured with a Keithley 2636B Source measure unit with an input impedance of $> 10^{14}$ Ω. Since the resistance was not always fully time stable, those have some uncertainty. We calculated the correction factor which needs to be multiplied to obtain proper voltages as $K = (R_{DC} + R_{MFLI})/R_{MFLI}$, utilizing the measured DC resistance between the input tips and the input resistance of the MFLI. To account for non-ohmic behavior, we corrected each datapoint for the fitted slope of the recorded IV curve between it and the previous datapoint. As this correction only fully applies in the high-field area between the input and not the full path to the output, it constitutes a slight underestimation of the output resistance.

Ohmic heating for the sequentially doped sample was calculated by the finite element model, described in section 3, to be excessively high and would be unstable in reality. In reality, upon heating the dopant rapidly sublimates, making the heating process self-limiting. [1] We therefore predict that the local conductance between the measurement tips is significantly lower than the measured global conductance between the contacts. This also explains why the conductance of the sequentially doped films shows little change with the application of field. The global conductance is too high to require any corrections but it is unclear how much ohmic heating and local resistance distort the result, making it hard to evaluate. Nonetheless fitting values are provided in Table S1.

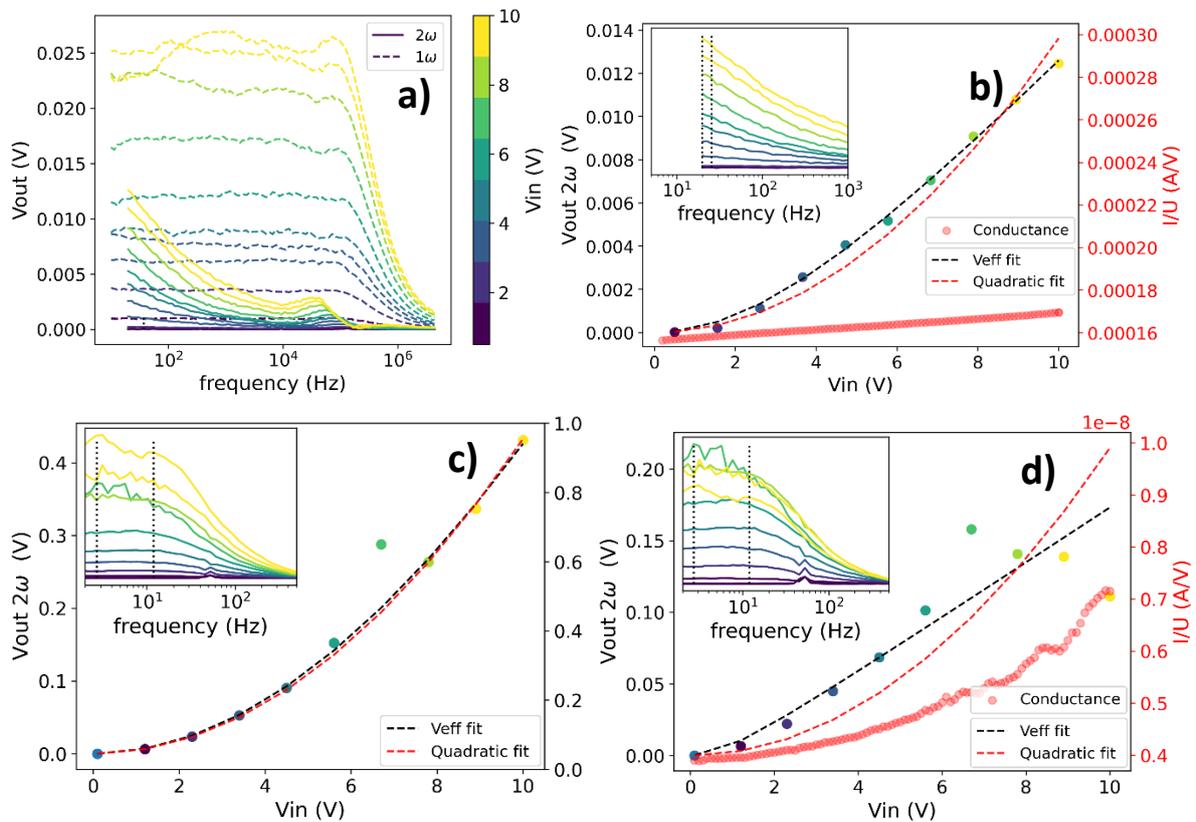

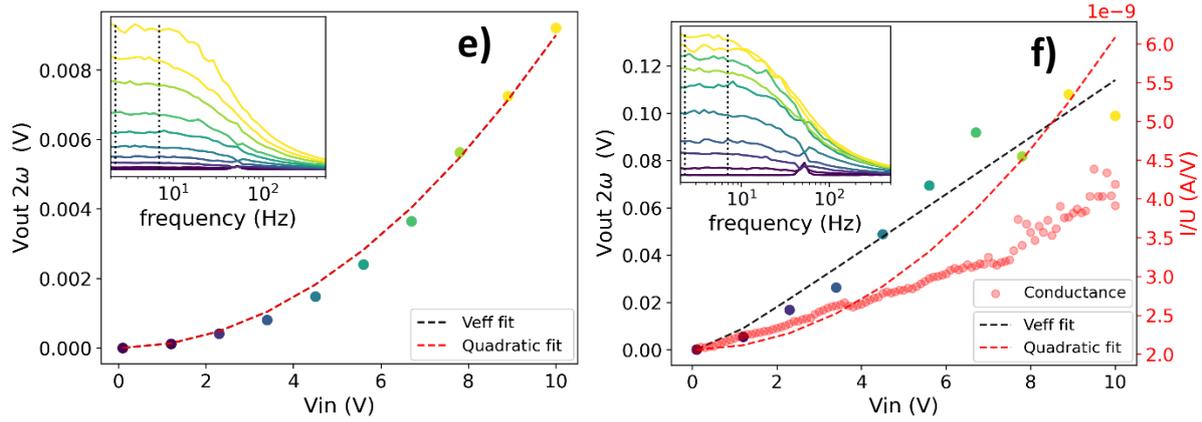

Figure S1: a) Depiction of the first and second harmonic signal in a sequentially doped sample. We suspect that the local resistance affecting the second harmonic signal is higher than the global resistance due to ohmic heating and subsequent sublimation of the dopant. The different phase behavior would then be caused by differing cutoff frequencies for the two signals. b-f) Cuts through the measurement for different films fitted with main text equation 2 and a quadratic function. Each point is the average between all measurements in the frequency range indicated by the black dotted lines. Plots featuring conductance data were corrected for the resistance mismatch between sample and measurement device. The P3HT films were sequentially doped (a,b) or 1% (c,d) to 0.01% (e,f) bulk doped with F4TCNQ. The corresponding parameters can be found in Table S1. Corresponding IV-curves measured between the input tips are shown and discussed in SI section 7.

| Doping | R ($\Omega$) | d/h(nm) | K | $\alpha/d_{fit}(10^{-3})$ | $S_{fit}$ (mV/K) | $C_{Om}$ | RR% |
|---|---|---|---|---|---|---|---|
| sequential | $6.3 - 5.5\,K$ | 35/60 | 1 | $5.9 \pm 0.6$ | $0.033 \pm 0.005$ | $1.3 * 10^{-4}$ | 96.2 |
| bulk 1% | $60 - 250M$ | 15/120 | 1 | $1.8 \pm 0.5$ | $0.290 \pm 0.150$ | $1.7 * 10^{-4}$ | 95.2 |
|  |  |  | 26.4 | $1.8 \pm 0.5$ | $7.77 \pm 3.96$ | $4.3 * 10^{-3}$ |  |
|  |  |  | 26-7 | $26 \pm 16$ | $0.070 \pm 0.050$ | $2.1 * 10^{-3}$ |  |
| bulk 0.1% *) | $500 - 700M$ | 35/60 | 1 | $0.4 \pm 5.5$ | $7 \pm 202$ | $2.0 * 10^{-4}$ | 95.2 |
|  |  |  | 71.5 | $0.4 \pm 5.5$ | $470 \pm 10640$ | $1.5 * 10^{-2}$ |  |
|  |  |  | 72-50 | $15 \pm 5$ | $0.334 \pm 0.185$ | $1.2 * 10^{-2}$ |  |
| bulk 0.01% **) | $170 - 290M$ | 35/60 | 1 | $0.05 \pm 12$ | $200 \pm 91000$ | $9.0 * 10^{-5}$ | 95.2 |
|  |  |  | 46.0 | $0.05 \pm 12$ | $10075 \pm 6000000$ | $4.1 * 10^{-3}$ |  |
|  |  |  | 46-10 | $48 \pm 66$ | $0.025 \pm 0.037$ | $1.3 * 10^{-3}$ |  |

Table S1: Comparison of key metrics for differently doped films. These include the doping ratio, the fitted Seebeck pre-factor $S_{fit}$ and fitted localization length over input distance $\alpha/d_{fit}$. Input distance d and measurement distance h are nominal values from the lithography process. R indicates the range of fitted input resistances and C Om the quadratic fit coefficient. K represents the range of correction factors multiplied to the output voltage before fitting. RR% indicates the degree of regioregularity of the utilized P3HT. *) The 0.1% bulk doped film measurement was from 0.1 to 2 volt in contrast to the other films measured to 10V. **) The structures were cleaned by oxygen plasma etching from a previous film before coating.

## 2. Setup checks:

As a sanity check we also investigated the second harmonic signal generated by our measurement setup. To this end we measured the response of an array of conventional resistors arranged in a T-shape to represent the experiment. As visible in figure S2 this system shows negligibly low second harmonic signals, one to two orders of magnitude below the raw data in Figure 2 and 3 of the main text, excluding the combination of function generator and lock-in amplifier as cause for the measured $2\omega$ signals.

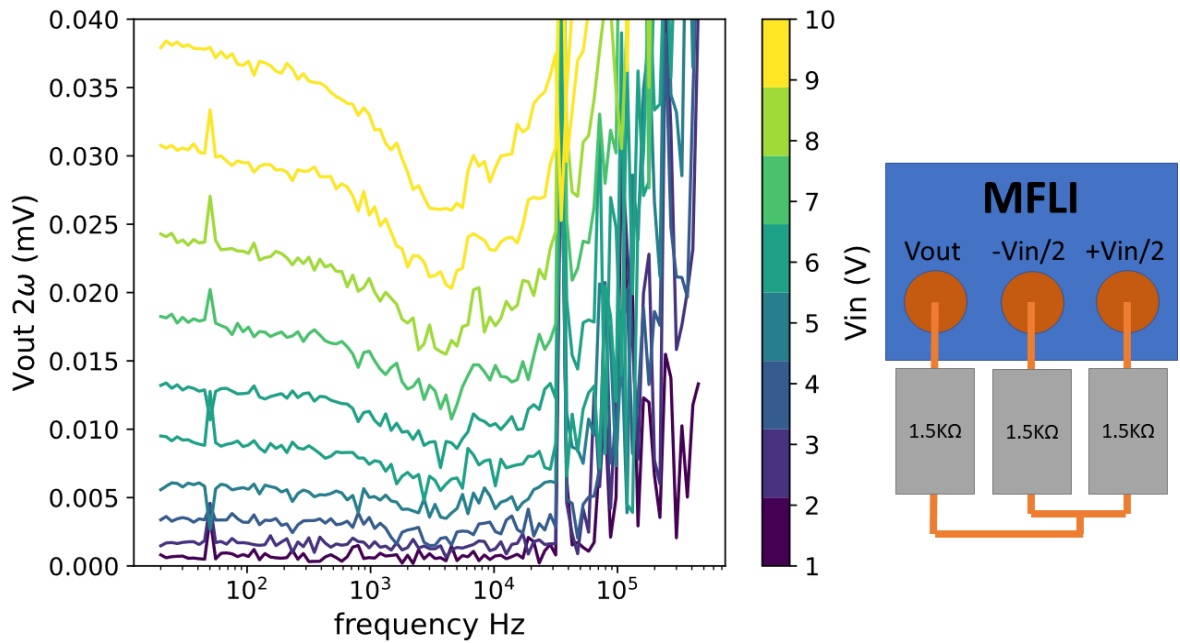

Figure S2: Response of 3 1.5 KΩ resistors connected to each other in a T-shape. Two ends were connected as inputs and one as output to the measurement device as depicted on the side.

Additionally, we also measured the response of an uncovered structure as shown in figure S3. A scanning electron microscopy picture is visible in figure 1 of the main text. While overall signal levels are expectedly low, the high frequency second harmonic data shows a resonance-like, input voltage dependent response. As this feature is not visible in the measured resistors we conclude that it is a feature of the measurement structure. Note also the absence of the plateau-shaped signal in the low-frequency (<1 kHz) regime that is present in P3HT-coated devices and that we interpret as due to the effective temperature-induced Seebeck effect.

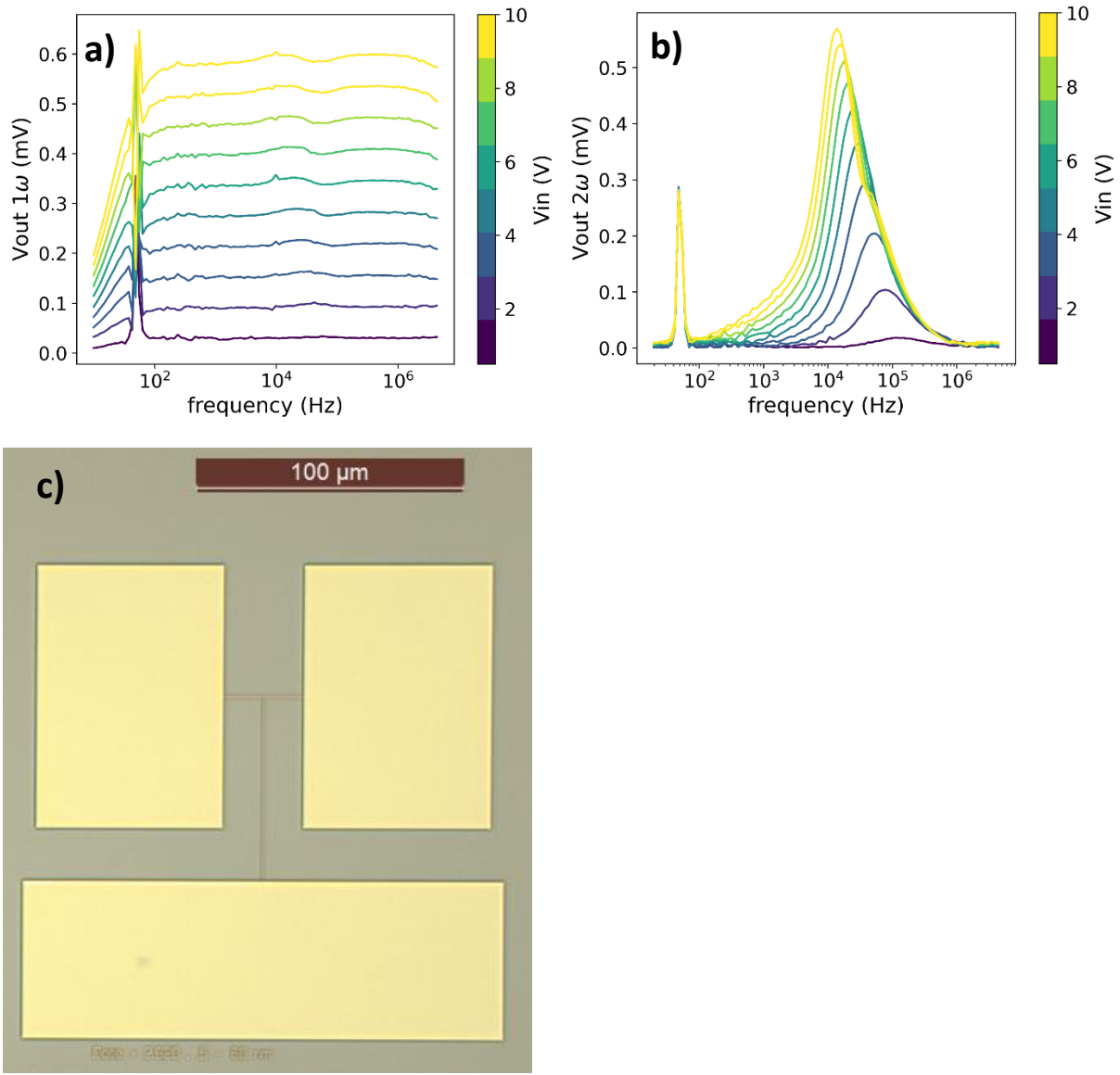

Figure S3: 1 omega a) and 2 omega b) output voltage of structures not covered with an organic layer. The input tip distance is 10 nm and the output tip distance is 60 nm. c) Microscope picture of an uncovered sample after liftoff.

The results in figure S3, and those in figure 2 of the main text, also allow us to rule out field emission of electrons from the tips. On P3HT-coated devices, the electrons would not be emitted into vacuum but into a solid were their mean free path would be minimal. Uncoated samples should therefore show more signal which they clearly do not. Furthermore, in such a scenario, the resistance of the sample would not be doping- but merely geometry-dependent. Finally, the voltage dependence would be of a very different, exponential instead of algebraic, form.

## 3. Finite element simulations:

In order to exclude ohmic heating we performed finite element simulations using the COMSOL 5.3 package of the worst-case scenario for heat generation by the applied signal. For simplicity and due to meshing limitations, we did not simulate the larger contacts structures (contact pads and lead wires to the tips) and only considered DC voltages. To reproduce the utilized structure, the tips were comprised of 5 nm of chromium and 10 nm of gold utilizing the standard COMSOL parameters for these materials. They had an input separation of 20 nm and the output tip was sitting at 60 nanometer distance from the center line. They are embedded in a 100-nanometer organic film with varying parameters as given in Table S. As many parameters were unknown, we made estimates based on literature data for similar materials. The upper film edge was set as a thermal insulator. The substrate was simulated as a 200 nm thick 320 x 420 nm silicon oxide slab with the temperature held at constant room temperature as bottom and side boundary conditions.

| Sigma S/m | Doping | Thermal conductivity W/(m K) | Heat capacity J/(Kg K) | Density Kg/m^3 | Tmax |
|---|---|---|---|---|---|
| 80 | Sequential | 0.3 | 1246 | 1150 | 2830 |
| 80 | | 1 | 1246 | 1150 | 1210 |
| 0.16 | >10% | 0.3 | 1246 | 1150 | 298.3 |
| 0.016 | ~0.01% | 0.3 | 1246 | 1150 | 293.66 |
| 0.008 | <=0.001% | 0.3 | 1246 | 1150 | 293.4 |

Table S2: Organic film parameters and results of the finite element simulations for the utilized measurement tips. The corresponding doping levels are estimates based on resistivity measurements performed in our group and given as molecular ratios.

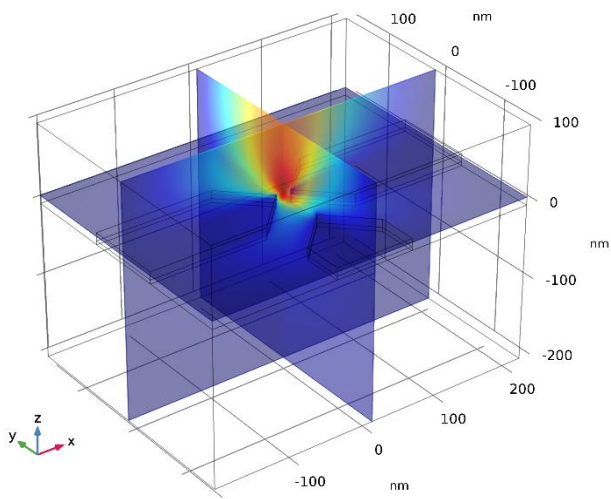

Figure S4: Finite element simulation of ohmic heating around the input tips. The difference between hot and cold (red and blue) regions depends on the doping level as described in Table S.

## 4. Stability and inter sample variation:

In our experiment output voltages decrease gradually over time, indicating that the doping level in our devices is not time stable. Even if this is a common observation for F4TCNQ-doped polymers, it makes comparing doping levels in different devices somewhat difficult and even hinders the comparison of different measurements on the same film. Surprisingly, this effect seems to outweigh the influence of tip separations in the experiment. In Figure S tips with different input distances have been measured in sequence on the same substrate. As visible, ordering the same measurements by time gives a much clearer trend than ordering them by input distance. The measurement where taken over the course of three days. We expect sublimation of the dopant to be responsible. [2] The differences are however not sufficient in order to fully explain the absence of input distance scaling.

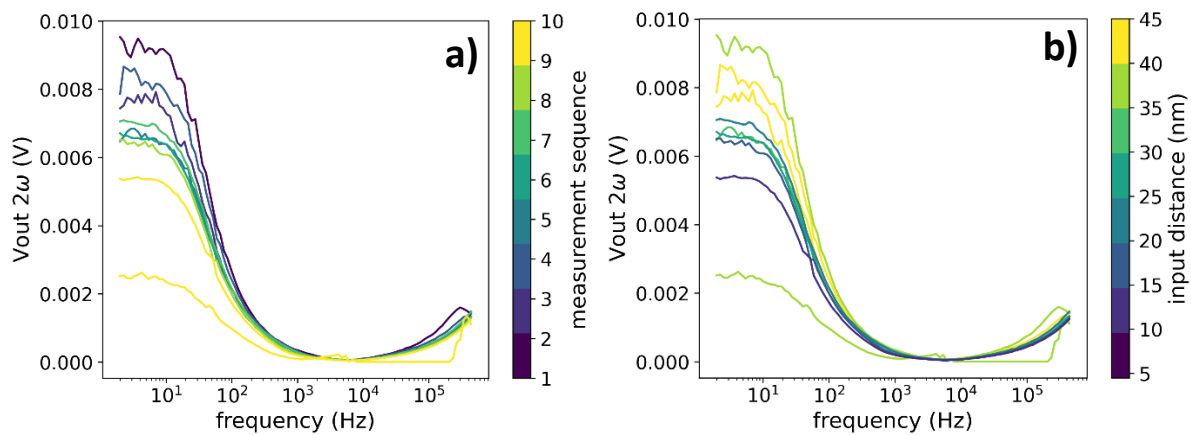

Figure S5: Measurements on different tips situated on the same substrate covered by a 0.001% bulk doped P3HT film with a 10 V input signal. Both graphs show the same measurements with different colors. Colors are ordered by either the time sequence in which the measurements were taken a) or by the input tip separation b). The measurement output distance was 60 nm.

We attribute this unexpectedly weak distance scaling to the small probed volume relative to the typical aggregate size in the used regioregular P3HT. [1,3,4] In contrast to conventional bulk devices, the active area here is expected to consist of at most a few aggregates making the experiment sensitive to their exact positioning. The transport through such a system is considered to be dominated by high conductance paths. The weak distance dependence visible in figure S5 b) could indicate that the dominating current pathways might be longer than the inter-tip distances we probed. This could also explain why we see diverging amplitudes across samples on the same substrate. Still, apart from the amplitude, the general functional shape is very consistent across different structures, indicating that the same mechanisms are at work irrespective of exact local morphology.

Similarly, also time stability can be affected by size. The active area is dominated by very few dopants. Given the molar mass of 168.3 g/mol as well as the density of 0.936 g/cm$^3$ the doping concentration of 0.1% would give us one dopant per 3000 nm$^3$. Therefore, the sample is expected to be more sensitive to the movement of individual dopants or changes in conformation than a bulk sample were these effects would average out. [5,6] These might be triggered thermally or by the applied field. This is consistent with occasional jumps we witness in our experiments as visible in figure S1 c).

This can also explain the relatively strong variation in the field dependent conductance measurements. Field-activated changes would be consistent with increasing measurement noise at high voltages. The resistance fits used to correct our data for impedance mismatch have errors in the percent range at high voltage. This causes the scattered data visible in figure S1 d) and f) when comparing to the uncorrected counterparts. Since lower input voltages seem to be less affected, this should be less of an issue in figure 3 b) (main text) and it is clearly not present in the more conductive sample shown in figure S1 a). These issues are not simply resolved by higher doping percentages. Additional bulk doping had a high tendency to form agglomerates and is therefore locally quite ill-defined, while sequential doping tends to give too high conductance as discussed above.

## 5. Drift diffusion simulations of field distribution:

In order to estimate how much the local geometry would distort our results we utilized a in house drift diffusion software to simulate the measurement tips. The simulation was performed in two steps. the first step estimated the field distribution. In a second step the local grid temperature was set to the effective temperature that would be caused by the predicted field strength and no fields were applied. The currents and voltages due to the temperature could now be read out. We compared the effective temperature difference between maximum field point and output, times the Seebeck coefficient, to the obtained voltages from the entire geometry. This way we could test if the maximum field is the main contribution to the measured output voltage. The obtained ratios ranged between 0.12 to 2.5, depending on sample geometry, with typical values lying close to unity. Hence, as the deviation is quite small in most cases, we do not account for it in fitting.

## 6. Phase behavior

Figure S shows a graph of the most common phase behavior in the measured samples. While low voltages below one volt show high scatter indicating that no clear phase was picked up higher voltages show very clear trends. High frequencies are dominated by the resonance also visible in the voltage amplitude of figure 2 (main text).

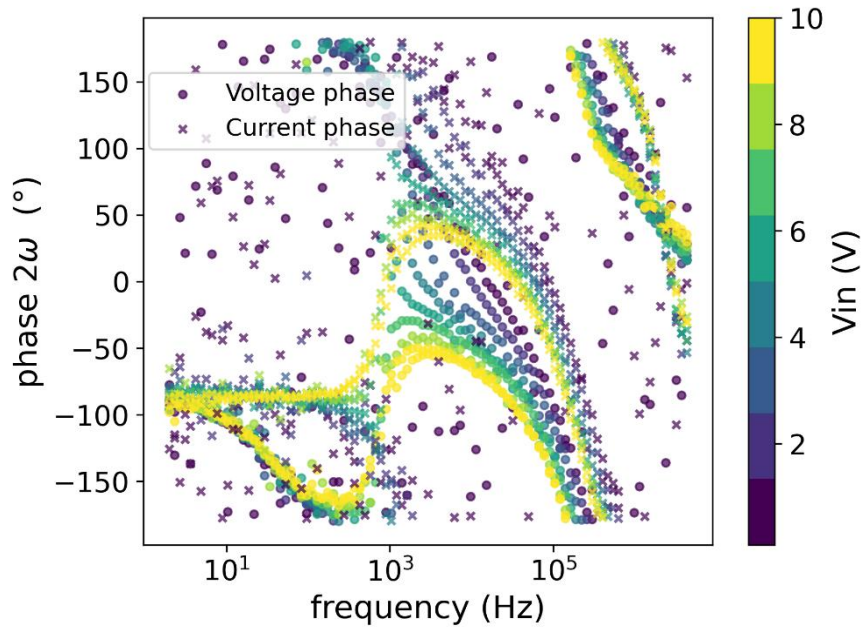

Figure S6: Voltage and current phase relative to the input voltage for a 1% molecular ratio bulk doped film. Input and output distances were 25 and 60 nm.

## 7. Current-voltage characteristics

We present here the IV-traces utilized for determining the conductance in Fig. 3b of the main text and SI Figs. S1 b,d,f. The corresponding IVs were measured utilizing a Keithley 2636B Source-Measure Unit. The curves were recorded with a minimum averaging time of 10 power line cycles. Nonlinearities are consistent with an increase in conductance due to increasing effective temperature. This behavior is for instance described by Nebel et al. [7].

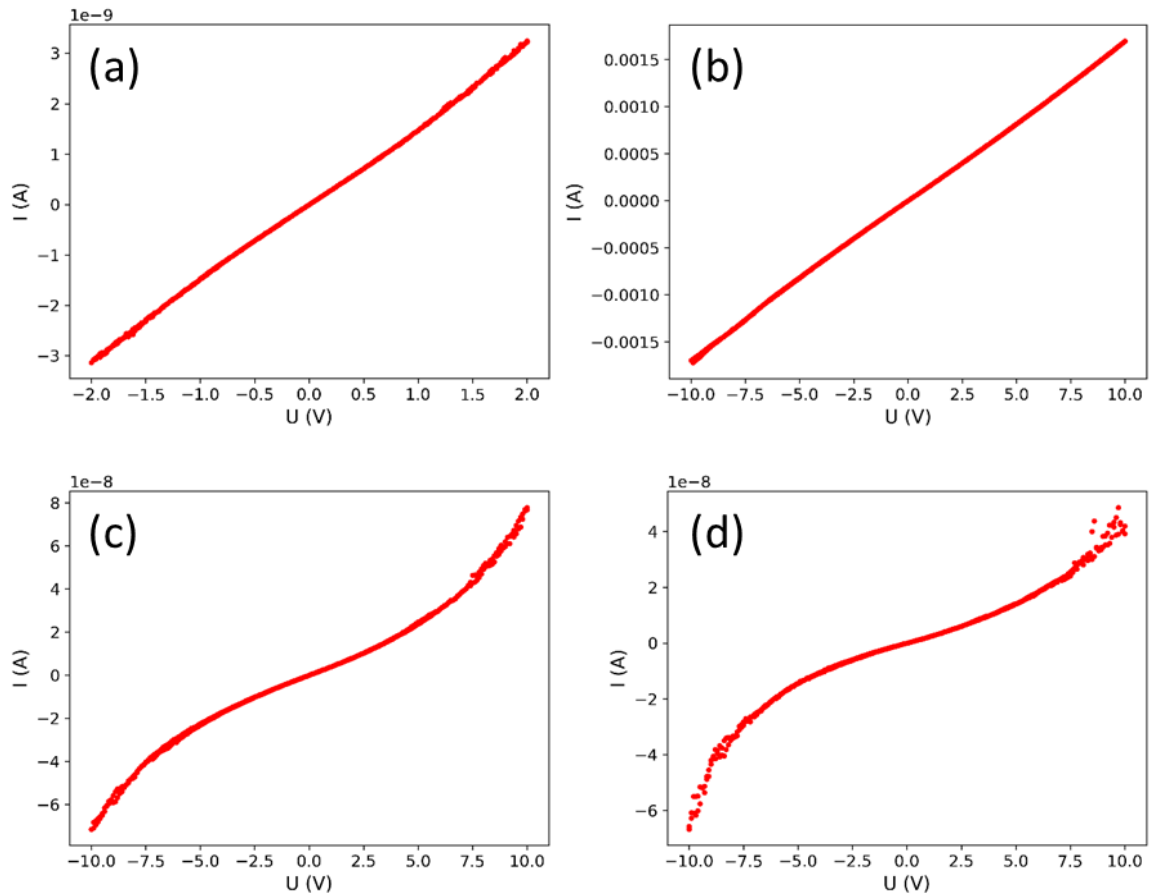

Figure S7: IV trace between the input tips of measurement structures corresponding to figures in the manuscript a) Fig. 3b; b) Fig. S1b; c) Fig. S1d and d) Fig. S1f.

While the nonlinearity of the IV-curves discussed above would only give rise to odd terms and thus not cause any second harmonic signals, any significant asymmetry between I(-V) and I(+V), i.e., rectification, would. Hence, we performed explicit calculations to check for higher order even terms in the IV-curves that could quantitatively explain our second harmonic signals and found this was generally not the case.